\documentclass[aip,apl, twocolumn, amssymb, amsmath, reprint]{revtex4}
\usepackage{mathrsfs}
\usepackage{graphicx}
\usepackage{amssymb}
\usepackage{dcolumn}
\usepackage{bm}
\usepackage{txfonts,epstopdf}
\usepackage{color}
\usepackage[colorlinks=true, letterpaper=true, pdfstartview=FitV, linkcolor=blue, citecolor=blue, urlcolor=blue]{hyperref}

\begin{document}

\title{All-electric spin transistor based on a side-gate-modulated two-dimensional topological insulator}

\author{Xianbo Xiao$^{1}$, Ying Liu$^{2}$}
\author{Zhengfang Liu$^{3}$, Guoping Ai$^{1}$}
\author{Shengyuan A. Yang$^{2}$}\email{shengyuan\_yang@sutd.edu.sg}
\author{Guanghui Zhou$^{4}$}
\email{ghzhou@hunnu.edu.cn}

\affiliation{$^1$School of Computer Science, Jiangxi University of
Traditional Chinese Medicine, Nanchang 330004, China}

\affiliation{$^2$Research Laboratory for Quantum Materials, Singapore University of Technology and Design, Singapore 487372, Singapore}

\affiliation{$^3$School of Basic Science, East China Jiaotong
University, Nanchang 330013, China.}

\affiliation{$^4$Department of Physics and Key Laboratory for
Low-Dimensional Quantum Structures and Manipulation (Ministry of
Education), Hunan Normal University, Changsha 410081, China}

\begin{abstract}
We propose and investigate a spin transistor device consisting of two ferromagnetic leads connected by a two-dimensional topological insulator as the channel material. It exploits the unique features of the topological spin-helical edge states, such that the injected carriers with a non-collinear spin-direction would travel through both edges and show interference effect. The conductance of the device can be controlled in a simple and all-electric manner by a side-gate voltage, which effectively rotates the spin-polarization of the carrier. At low voltages, the rotation angle is linear in the gate voltage, and the device can function as a good spin-polarization rotator by replacing the drain electrode with a paramagnetic material.
\end{abstract}

\maketitle

The discovery of topological insulators (TIs) has generated enormous interest in exploring their fundamentally new physical properties as well as possible technological applications.\cite{review1,review2} Particularly, the two-dimensional (2D) TI, also known as the quantum spin Hall insulator (QSHI), possesses a pair of topological transport channels confined at the sample edge while the bulk is insulating.\cite{Kane2005,BHZ} These channels are spin-polarized and helical, i.e., opposite spin states counter-propagate at a given edge. Furthermore, they are protected against backscattering from non-magnetic impurities, hence are in-principle dissipationless. Due to these unique features, QSHIs hold great promise for spintronics applications. However, from a device point of view, the robustness of the edge channels also poses challenge for designing efficient and effective control methods. In previous studies of QSHI-based devices, several schemes for controlling the edge-channel transport have been proposed, which are mainly based on quantum point contact (or nano-constriction) structures which couple the channels on opposite edges,\cite{Krue2011,Dolc2011,Liu2011,Li2012,Rome2012,Zhan2011,Dolc2013,Fu2014,Chen2014} or by using external magnetic field or exchange field.\cite{Cheng2014,llan2012,Soor2012,Zhan2012,Maci2010,Hofe2014}

\begin{figure}
\includegraphics[width=9cm]{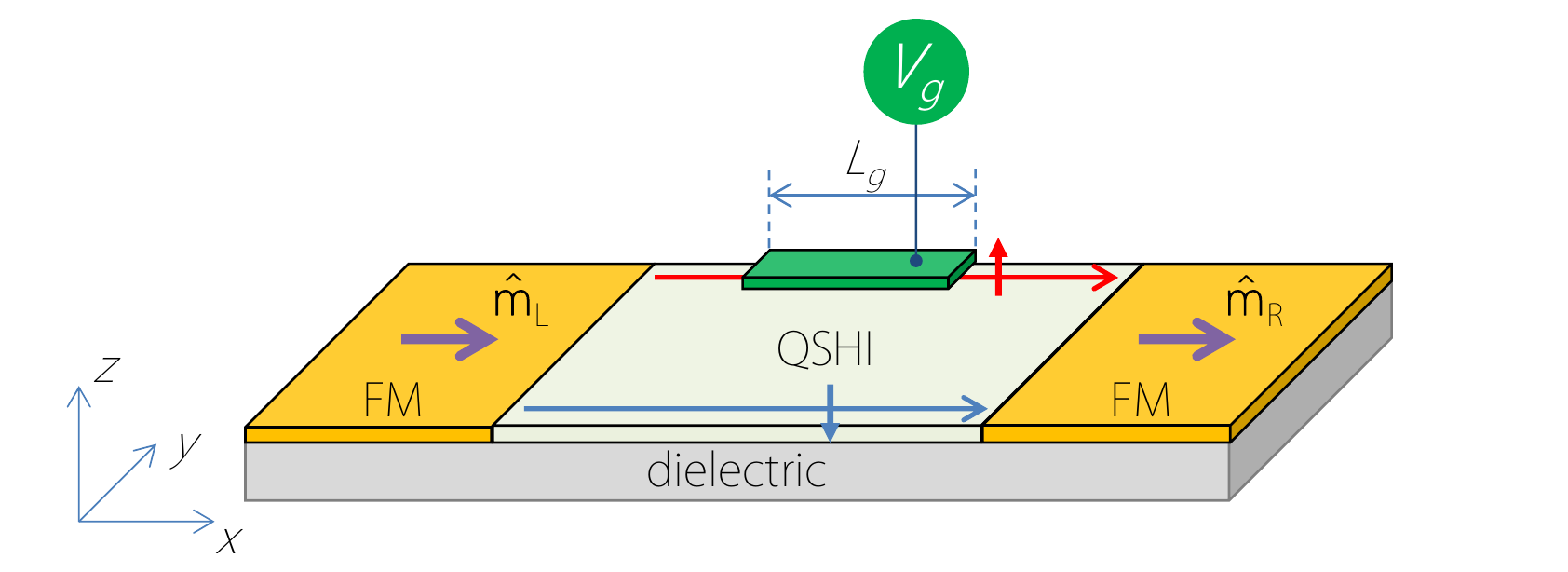}
\caption{Schematic figure showing the QSHI-based spin transistor. It consists of two ferromagnetic leads and a QSHI channel in which the transport is through the topological spin-helical edge channels. Here only the edge channels propagating from left to right are shown. The transport is controlled by a side gate at one of the edges. }
\label{fig1}
\end{figure}

Motivated by the rapid progress in the field of topological insulators and by the experimental advance in fabricating high-quality QSHI quantum well structures which could achieve ballistic transport over micron-scale,\cite{Konig,Brun2010,Knez2011,Brun2012,Knez2012} in this work, we propose and study a new QSHI-based spin transistor device (see Fig.~\ref{fig1}). Different from the existing proposals, the design exploits the spin-polarized transport at both edges of the QSHI channel material, and relies on the interference between them which can be effectively controlled by a small side-gate voltage. It has the merits of simplicity, low power-consumption, and all-electric controllability. Moreover, the device could function as a good spin-polarization rotator with slight modification of the design, and it can also be used to probe the edge states of a QSHI.

As schematically shown in Fig.~\ref{fig1}, the device consists of two ferromagnetic electrodes connected by a QSHI channel material. A side-gate is deposited along one of the sample edge (here chosen as the upper edge) and is used to tune the local chemical potential around the edge. The operation of the device only relies on the general features of QSHI state, not depend on the specific material. Here, for concreteness, we take HgTe quantum well as the channel material, which has been experimentally demonstrated as a QSHI.\cite{Konig} Its low-energy physics is described by the four-band Bernevig-Hughes-Zhang model around the $\Gamma$-point of the Brillouin zone.\cite{BHZ} Written in the basis of $|E_1,+\rangle$, $|H_1,+\rangle$, $|E_1,-\rangle$, $|H_1,-\rangle$, where $E_1$ and $H_1$ denote the electron and the heavy-hole subbands, and $\pm $ represent spin-up and spin-down states along the $z$-axis, the Hamiltonian takes the form
\begin{equation}\label{BHZ}
H_0=\varepsilon(\bm k)+M(\bm k)\tau_z+A(k_x\sigma_z\tau_x-k_y\tau_y).
\end{equation}
Here $\bm k=(k_x,k_y)$ is the 2D wave-vector, $\varepsilon(\bm k)=C-Dk^2$, $M(\bm k)=M_0-Bk^2$, the coefficients $A$, $B$, $C$, $D$, and $M_0$ are material-dependent parameters, the Pauli matrices $\tau$ denote the $E_1$ and $H_1$ states and $\sigma$ represent the spin states. We model the channel of device with width $W$ and length $L$ by discretizing Eq. (\ref{BHZ}) onto a 2D square lattice.\cite{Jiang} The effect of the side-gate is modelled by adding an on-site potential $V_g$ to a region with dimension $L_g\times W_g$ around the upper edge.

The two electrodes can in general be any ferromagnetic metals, and the essential physics that we discuss below does not depend on the specific material. Here, in order to minimize the contact resistance, we model the leads using the same $H_0$ in Eq. (\ref{BHZ}) with an additional exchange coupling term. Experimentally, this can be realized by Mn-doping in HgTe quantum wells.\cite{Furd1988,Beck2001} The resulting exchange term can be written as\cite{Liu,Liu2013} $H^{m}_\text{L(R)}=g_e\bm{m}_{\text{L(R)}}\cdot\bm \sigma(1+\tau_z)/2+g_h\bm{m}_{\text{L(R)}}\cdot\bm \sigma(1-\tau_z)/2$, where $g_{e(h)}$ is the $g$-factor for the $|E_1\rangle$ ($|H_1\rangle$) orbital, $\text{L(R)}$ stands for the left (right) lead, and $\bm m$ is a vector along the exchange field whose direction can be controlled experimentally, e.g., by an external magnetic field, and its strength can be tuned, e.g., by the doping concentration.

\begin{figure}
\includegraphics[width=9cm]{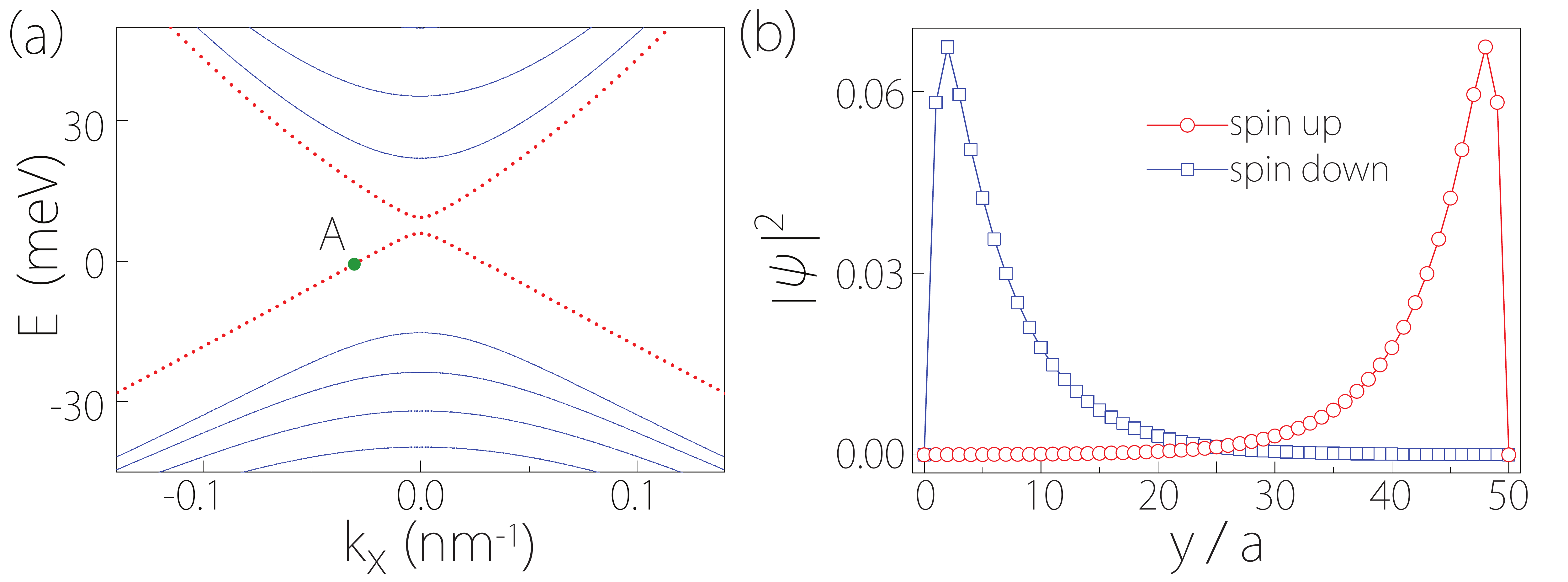}
\caption{(a) Energy spectrum of an extended QSHI ribbon with a width of $50a$ ($a=2.0$ nm), showing topological edge states (red dotted lines) inside the gap of bulk bands (blue solid lines). (b) Spatial distribution of the two spin-polarized edge states marked by A (at energy -1.25 meV) in (a)). The spin-up channel is localized at the upper edge and the spin-down channel is at the lower edge. The parameter values are taken for HgTe quantum wells: $A=365$ nm$\cdot$meV, $B=-700$ nm$^2\cdot$meV, $D=-500$ nm$^2\cdot$meV, $M_0=-10$ meV, and the constant energy shift $C$ is taken to be zero.}
\label{fig2}
\end{figure}

The transport property of the device is studied using the Landauer-B\"{u}ttiker formalism combined with non-equilibrium Green's function techniques. The two-terminal conductance for the considered system is given by\cite{Datta}
\begin{eqnarray}\label{GG}
G=\frac{e^2}{h}\mathrm{Tr}[\Gamma_{\mathrm{L}}G^{\mathrm{r}}\Gamma_{\mathrm{R}}G^{\mathrm{a}}],
\end{eqnarray}
where $e$ is the electron charge and $h$ is the Planck's constant, $G^{r/a}$ is the retarded/advanced Green's functions of the channel region,  $\Gamma_\text{L(R)}$ are the linewidth functions describing the coupling between the left (right) lead with the central channel region, and the trace is taken over both the spatial and spin degrees of freedom. The linewidth functions can be obtained from $\Gamma_\alpha=i(\Sigma_\alpha^r-\Sigma_\alpha^a)$, where $\Sigma_\alpha^{r/a}$ is the retarded/advanced self-energies for the $\alpha$ lead ($\alpha=\text{L, R}$). All the quantities in Eq. (\ref{GG}) are evaluated at the Fermi level. In the modeling, we assume two leads to be semi-infinite and have the same width $W$ as the channel region. The self-energies of the leads as well as $G^{r/a}$ can be computed using a recursive method.\cite{Ando}

We first consider the configuration in which the two leads are spin-polarized along the $x$-direction, i.e., $\hat{\bm m}_{\text{L,R}}=\hat{x}$, where the hat indicates a unit vector. For the two leads, to make the physical picture more transparent, we choose the chemical potential such that the carriers on the Fermi level are fully spin-polarized, i.e., the source and drain correspond to perfect spin polarizer and analyzer, respectively.
In the central channel region, the chemical potential is required to be in the bulk bandgap such that the system is in a QSHI state. Figure~\ref{fig2} shows the electronic band structure of a corresponding infinitely-long (along $x$) QSHI ribbon with $W=50a$, where $a=2.0$ nm is the lattice constant of the square lattice used for our numerical calculation. One can observe the topological spin-helical edge states in the bulk bandgap, marked by the red dotted lines. In fact, each line is doubly degenerate because there is one state from each edge. At a given edge, there is a pair of counter-propagating states with opposite spin-polarization along the $z$-axis. If we consider the modes going from source (left) to drain (right), then there are two channels: one spin-up ($|z;+\rangle$) channel at upper edge and one spin-down ($|z;-\rangle$) channel at lower edge, as indicated in Fig.~\ref{fig1}. Figure~\ref{fig2}(b) shows the spatial distribution of the two modes marked by point A in Fig.~\ref{fig2}(a), which verifies the above picture. In Fig.~\ref{fig2}(a), one observes a small gap opened in the edge-states spectrum. This is due to the finite width of the sample such that the states on the two edges can have a small hybridization.\cite{Zhou2008} Several previous proposals tried to use this inter-edge hybridization to control the edge-state transport.\cite{Krue2011,Dolc2011,Liu2011,Li2012,Rome2012,Zhan2011,Dolc2013,Fu2014,Chen2014} In contrast, the device operation here does not rely on this mechanism, and in our following discussion the Fermi level always lies outside of this hybridization gap.

Now for an incoming electron from the left lead, it is spin-polarized along the $x$-direction, i.e., with spin state $|x;+\rangle$. After it enters the channel region, it can be transmitted by coherently splitting and propagating through both edges, with $|x;+\rangle=\frac{1}{\sqrt{2}}(|z;+\rangle+ |z;-\rangle)$. Indeed, the superposition of the two edge states offers a single channel for a spin-polarized carrier. Due to the reflections at the entrance and exit of the channel region, there will be the usual Fabry-Perot type oscillations in the conductance as a function of the channel length. In the following analysis, we fix the channel length such that resonant transmission occurs at $V_g=0$. (Note that the modulation effect by $V_g$ that we discuss below does not depend on the choice of channel length, which simply amounts to a nonzero offset in $V_g$.) Then by the above analysis, we expect to have a conductance close to $e^2/h$ for $V_g=0$.

\begin{figure}
\includegraphics[width=9.1cm]{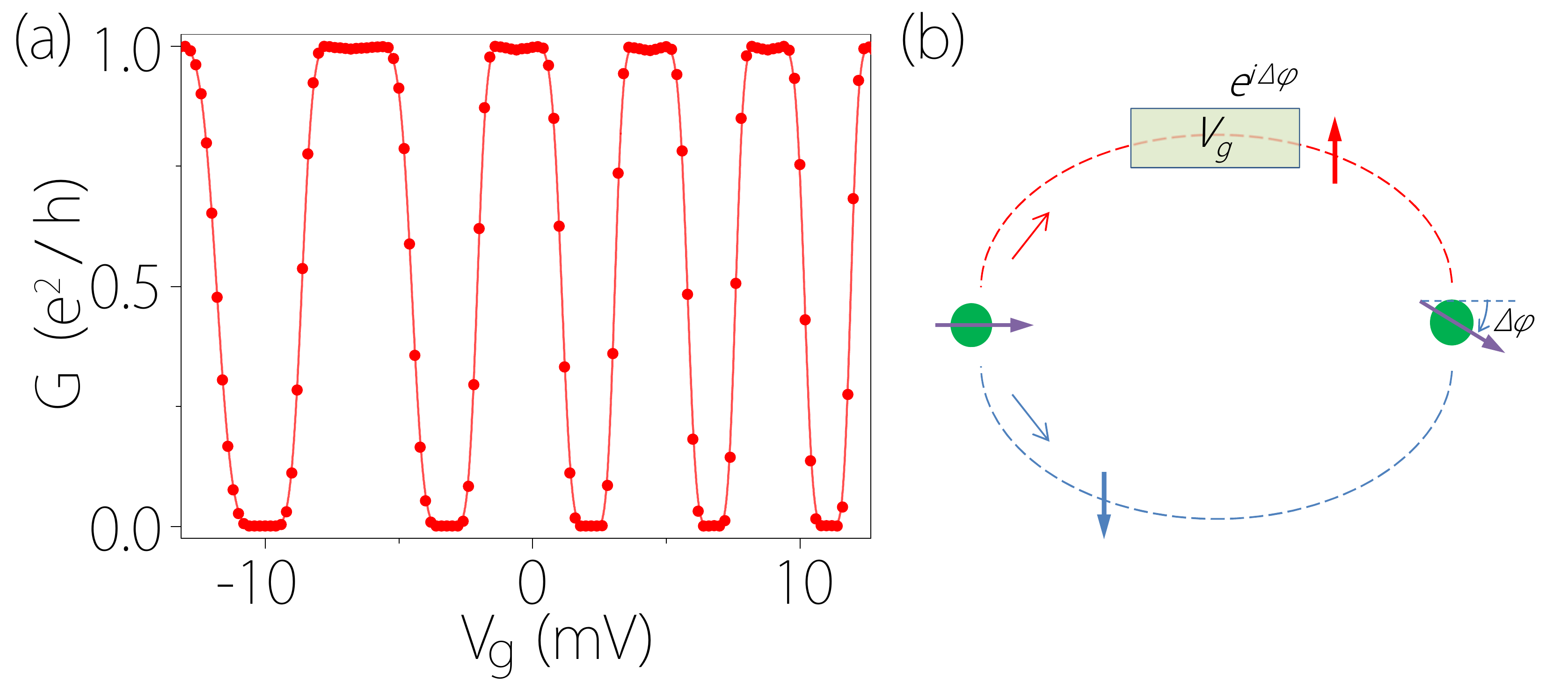}
\caption{(a) Conductance versus the side gate voltage $V_g$. The model parameters are the same as in Fig.~\ref{fig2}, and $g_e=0.39$, $g_h=1.61$, $m_L=m_R=2.0$ meV, channel length is set to $285a$, $L_g=180a$, $W_g=10a$, and Fermi level $E_F=-1.25$ meV. (b) Schematic figure showing the working principles of the device. An incoming electron polarized along $\hat{x}$ get split into the two spin-polarized edge channels. The transmission through the upper edge is modulated by the side gate and acquires an extra phase. When the carrier reaches the drain, its polarized is rotated in the $xy$-plane by an angle $\Delta\varphi$. }
\label{fig3}
\end{figure}

By applying a gate voltage $V_g$, as shown in Fig.~\ref{fig3}(a), one observes that the conductance begins to drop when $V_g$ increases. It was almost completely suppressed at a relatively small gate voltage $\sim2.15$ mV, and after which it starts to increase. With increasing $V_g$, the conductance exhibits an interesting periodic modulation between 1 and 0 (here and hereafter, conductance is stated in units of $e^2/h$). For $V_g<0$, the result is quite similar, although the oscillation period is slightly increased due to deviation from perfect linear dispersion (Fig.~\ref{fig2}(a)).

\begin{figure}
\includegraphics[width=8cm]{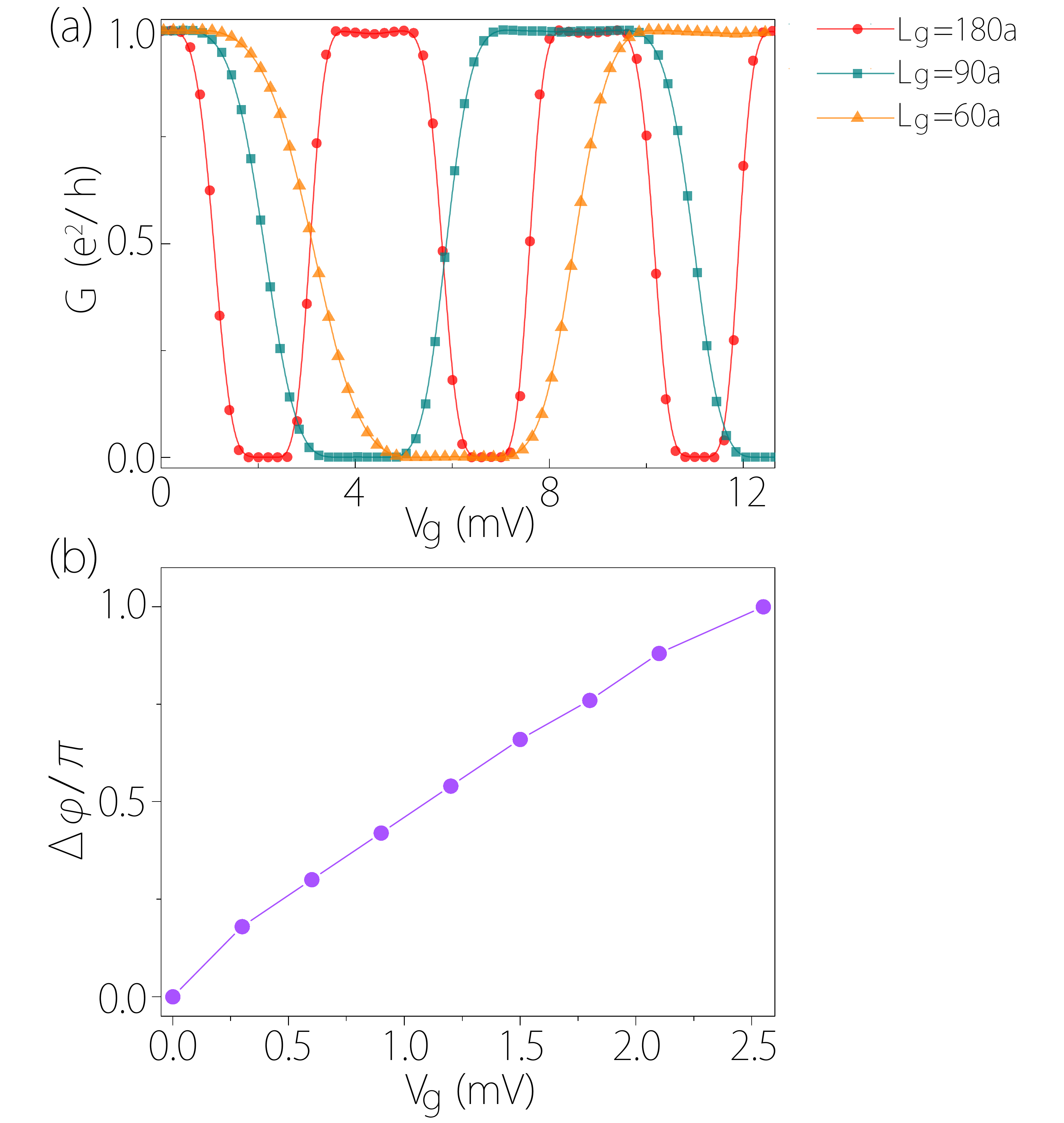}
\caption{(a) Modulation of conductance by $V_g$ for different side-gate length $L_g=180a$, $90a$, and $60a$. (b) The spin rotation angle $\Delta \varphi$ versus $V_g$ for small voltages and $L_g=180a$. The other parameters are the same as in Figs.~\ref{fig1} and \ref{fig2}.}
\label{fig4}
\end{figure}

To understand this phenomenon, we notice that $V_g$ only affects the upper edge. Its main effect is the modulation of the phase of carrier passing through the upper edge, which is for the $|z;+\rangle$ spin state. When $V_g$ is small and slowly varying, an extra phase is acquired for the transmission through the upper edge, which can be approximated as
\begin{equation}\label{phi}
\Delta\varphi=2\pi eV_gL_g/(hv_F),
\end{equation}
where $v_F$ is the group velocity of the edge state. As a result, when the carrier reaches the right lead, its state becomes $\frac{1}{\sqrt{2}}(e^{i\Delta\varphi}|z;+\rangle+|z;-\rangle)$, i.e., its polarization is rotated from $\hat{x}$ by an angle $(-\Delta\varphi)$ in the $xy$-plane. Hence the transmission probability and the conductance are reduced by a factor $\cos^2(\Delta\varphi)$. In Fig.~\ref{fig3}(a), if we take the point $V_g=2.15$ mV as corresponding to the condition that $\Delta\varphi=\pi$, then from the above approximation we can estimate the Fermi velocity $v_F\simeq 3.83\times 10^5$ m/s, which agrees well with the value $\simeq 3.82\times 10^5$ m/s extracted from the energy spectrum in Fig.~\ref{fig2}(a). When $V_g$ is large, the spatial distribution of the edge state will also be affected. For example, for a large negative voltage, the edge channel will be pushed away from the gated region. Then the dependence of $\Delta\varphi$ on $V_g$ for large voltages generally deviates from a simple linear relationship, which is also reflected in the shape change of the conductance curve for large $V_g$. Figure~\ref{fig4}(a) shows the conductance curves for different gate length $L_g$. The modulation period increases with decreasing $L_g$, and scales approximately linearly in $L_g$ as predicted in Eq. (\ref{phi}).

From the above analysis, we see that the device effectively rotates the spin-polarization of the incoming electron in the $xy$-plane by an angle controlled by $V_g$. For a given $V_g$ we can determine the spin rotation angle $\Delta\varphi$ by rotating $\hat{\bm m}_\text{R}$ in the $xy$-plane and search for the configuration at which $G=0$, which is effectively like making the ferromagnetic drain as an analyzer for the spin direction. Figure~\ref{fig4}(b) shows the resulting $\Delta\varphi$ vs. $V_g$ for small voltages, which indeed agrees well with our approximation in Eq. (\ref{phi}).

If we replace the drain electrode by a paramagnetic metal, the device can act as a spin-polarization rotator, which rotates the spin-polarization of the current, similar to an optical polarization rotator. In the original Datta-Das spin transistor,\cite{Datt1990} the spin-polarization of the carrier is rotated by gate-controlled Rashba spin-orbit coupling. However, the carriers entering the channel are moving along different directions, leading to a spread of rotation angles when they reach the drain, not to mention the spin-relaxation inside the channel.\cite{Zuti2004}
In comparison, here the edge states are fully spin-polarized, principally without spin-relaxation; and these are one-dimensional channels, hence angular spread should be much smaller. Consequently, the current spin transistor as well as the spin-polarization rotator could have a better performance than the traditional designs.

\begin{figure}
\includegraphics[width=8.8cm]{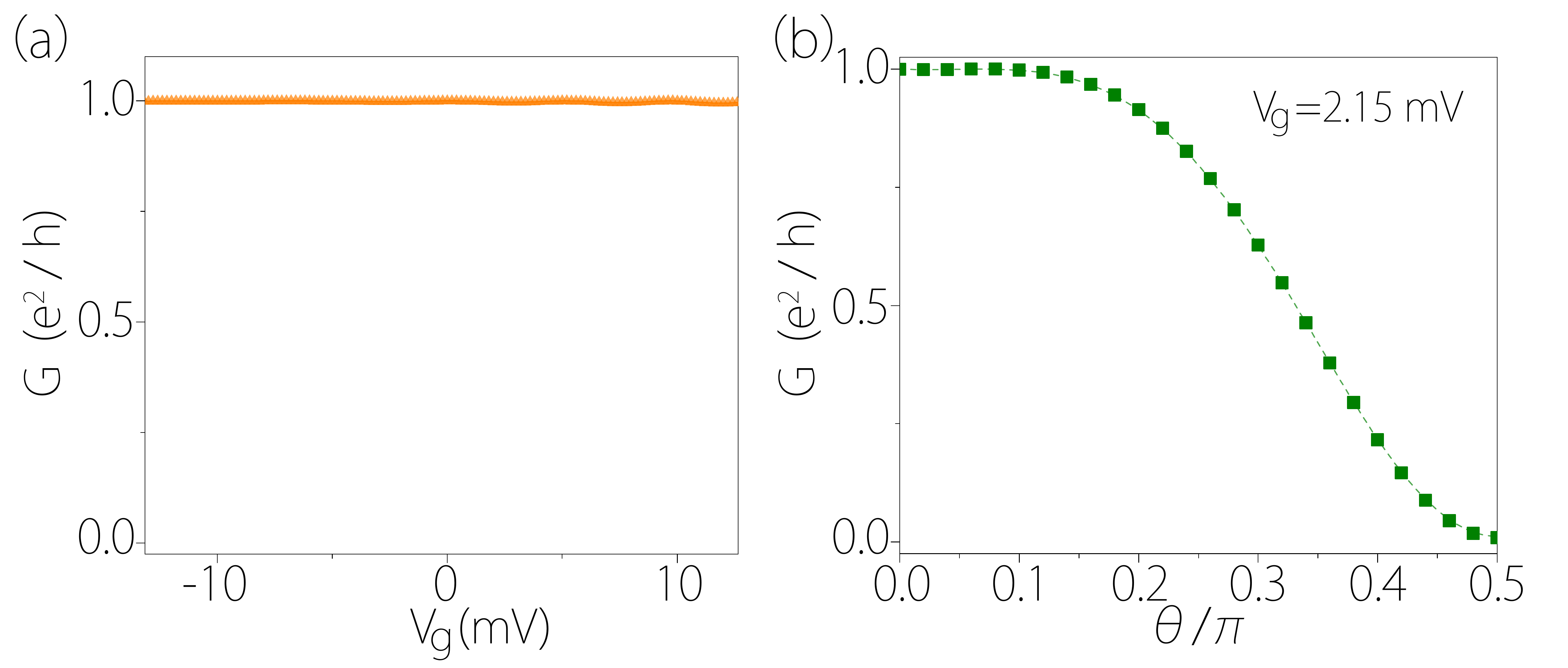}
\caption{(a) Conductance versus $V_g$ for both leads polarized along $\hat{z}$-direction. (b) Conductance as a function of the polar angle $\theta$
of $\hat{\bm m}_\text{L,R}$ (in the same direction) at a fixed voltage $V_g=2.15$ mV. The other parameters are the same as in Figs.~\ref{fig1} and \ref{fig2}.}
\label{fig5}
\end{figure}

The observed electric modulation of the conductance is a result of the interference between the transport at the two edges. It fully exploits the features of QSHIs: edge channels are spin-polarized and both edges conducts. It is crucial that both edges participate in the transport, which is the case when the leads are polarized in directions perpendicular to $\hat{z}$. If we let the leads polarized along $z$-direction, then the carriers will only travel along one of the edges, hence the modulation effect is expected to be suppressed. Figure~\ref{fig5}(a) shows the result for $\hat{\bm m}_\text{L,R}=\hat{z}$, which indeed confirms our expectation.

For the intermediate case with $0<\theta_{\text{L,R}}<\pi/2$, where $\theta_{\text{L(R)}}$ is the polar angle for $\hat{\bm m}_\text{L(R)}$ measured from $z$-axis, if $\hat{\bm m}_\text{L}$ and $\hat{\bm m}_\text{R}$ are aligned in the same direction (with angle $\theta$), then based on our above discussion, varying $\theta$ could tune the degree of interference between the two edges. Figure~\ref{fig5}(b) shows the conductance as a function of $\theta$ at $V_g=2.15$ mV which corresponds to the first $G=0$ point in Fig.~\ref{fig2}(a). One observes that $G$ is continuously tuned from $1$ to $0$ between $\theta=0$ where there is no interference and $\theta=\pi/2$ where there is strong destructive interference. In addition, for a given $\hat{\bm m}_\text{L}$, to maximized the on-off ratio of the device, one may choose $\hat{\bm m}_\text{R}$ such that $\theta_\text{R}=\pi-\theta_\text{L}$ and $\phi_\text{R}=\phi_\text{L}$ (or $\phi_\text{L}+\pi$), then the conductance would be modulated between 0 and $\sin^2\theta$.

Some remarks are in order here. As we mentioned before, several previous device proposals rely on the inter-edge coupling between the states on opposite edges, e.g., by designed constriction structures. In the contrary, here the decoupling between the two edges is desired. To this end, one needs the localization length (along the width) of the edge states $\xi$ to be much smaller than the sample width $W$. This can be achieved either by increasing $W$, or by choosing a channel material with a large bulk bandgap since it has been shown that $\xi$ is inversely proportional to the bulk gap.\cite{Zhou2008}

For a QSHI, its edge states are topologically-protected against non-magnetic scatterers.\cite{review1,review2,Takagaki} With preserved time reversal symmetry (in the channel region), the side-gate modulation effect should survive against elastic scatterings since the phase coherence is preserved (although there may be a constant phase shift at zero voltage). However, inelastic scatterings would randomize the phase hence suppress the modulation effect. Experimentally, it has been demonstrated that the electron mean free path over several microns can be achieved in HgTe quantum well structures at low temperatures.\cite{Konig,Brun2010,Brun2012} For a device with dimensions less than a micron, one can assume a ballistic transport and the effect should be observable.

Finally, from our analysis of the underlying physics, it should be clear that the specific choice of the material model here is not essential. Any QSHI connecting two ferromagnetic leads should exhibit similar physics. For other QSHIs, the spin-polarization of the edge channels may not be along the $z$-direction (for HgTe quantum wells, the polarization may also slightly deviate from $\hat{z}$ if possible structure inversion asymmetry is present\cite{Roth2010}). However, the edge channels would still form time-reversal pairs and have opposite spin-polarization along certain direction $\hat{n}$. Then the maximum modulation effect would occur when the lead polarization is perpendicular to $\hat{n}$. Using this effect, one can actually probe the spin-polarization of the QSHI edge states.

In summary, we have proposed and investigated a new QSHI-based spin transistor device. We demonstrate that its conductance can be effectively modulated between 0 and $e^2/h$ using a side-gate in an all-electric manner. We clarify the underlying physics, and show that by replacing the drain electrode with a paramagnetic metal, the device can also act as a good spin-polarization rotator. Based on a different working principle from other designs, the device has the advantages of low-dissipation, structural simplicity, and all-electric controllability. The switching can be achieved at a small applied voltage, and compared with Datta-Das type spin transistors, its performance does not suffer from the spread of spin precession angles.

The authors thank D.L. Deng for valuable discussions. This work was supported by NSFC (Grant Nos. 11264019, 11364019, 11464011, and 11274108),
by the Natural Science Foundation of Jiangxi (Grant No. 20151BAB202007),
and by SUTD-SRG-EPD2013062.


\begin{thebibliography}{}

\bibitem{review1} M. Z. Hasan and C. L. Kane, Rev. Mod. Phys. {\bf 82}, 3045 (2010).
\bibitem{review2} X.-L. Qi and S.-C. Zhang, Rev. Mod. Phys. {\bf 83}, 1057 (2011).

\bibitem{Kane2005} C. L. Kane and E. J. Mele, Phys. Rev. Lett. {\bf 95}, 146802 (2005).
\bibitem{BHZ} B. A. Bernevig, T. L. Hughes, and S.-C. Zhang, Science {\bf 314}, 1757 (2006).

\bibitem{Krue2011} V. Krueckl and K. Richter, Phys. Rev. Lett. {\bf 107}, 086803 (2011).
\bibitem{Dolc2011} F. Dolcini, Phys. Rev. B {\bf 83}, 165304 (2011).
\bibitem{Liu2011} G. Liu, G. Zhou, and Y.-H. Chen, Appl. Phys. Lett. {\bf 99}, 222111 (2011).
\bibitem{Li2012} Y. Li, M. B. A. Jalil, S. G. Tan, and G. Zhou, {\bf 112}, 063710 (2012).
\bibitem{Rome2012} F. Romeo et al., Phys. Rev. B {\bf 86}, 165418 (2012).
\bibitem{Zhan2011} L. B. Zhang, F. Cheng, F. Zhai, and K. Chang, Phys. Rev. B {\bf 83}, 081402 (2011).
\bibitem{Dolc2013} G. Dolcetto et al., Phys. Rev. B {\bf 87}, 085425 (2013).
\bibitem{Fu2014} H.-H. Fu, D.-D. Wu, and L. Gu, Appl. Phys. Lett. {\bf 116}, 064511 (2014).
\bibitem{Chen2014} W. Chen et al., Phys. Lett. A {\bf 378}, 1893 (2014).


\bibitem{llan2012} R. llan et al., Phys. Rev. Lett. {\bf 109}, 216602 (2012).
\bibitem{Soor2012} A. Soori, S. Das, and S. Rao, Phys. Rev. B {\bf 86}, 125312 (2012).
\bibitem{Zhan2012} Y.-T. Zhang, F. Zhai, Z. Qiao, and Q.-F. Sun, Phys. Rev. B {\bf 86}, 121403 (2012).
\bibitem{Maci2010} J. Maciejko, E.-A. Kim, and X.-L. Qi, Phys. Rev. B {\bf 82}, 195409 (2010).
\bibitem{Cheng2014} F. Cheng, L. Z. Lin, and D. Zhang, Solid State Commun. {\bf 188}, 45 (2014).
\bibitem{Hofe2014} P. P. Hofer et al., EPL {\bf 107}, 27003 (2014).

\bibitem{Konig} M. K\"{o}nig et al., Science \textbf{318}, 766 (2007).
\bibitem{Brun2010} C. Br\"{u}ne et al., Nat. Phys. {\bf 6}, 448 (2010).
\bibitem{Brun2012} C. Br\"{u}ne et al., Nat. Phys. {\bf 8}, 485 (2012).
\bibitem{Knez2011} I. Knez, R.-R. Du, and G. Sullivan, Phys. Rev. Lett. {\bf 107}, 136603 (2011).
\bibitem{Knez2012} I. Knez, R.-R. Du, and G. Sullivan, Phys. Rev. Lett. {\bf 109}, 186603 (2012).

\bibitem{Jiang} H. Jiang, L. Wang, Q.-F. Sun, and X. C. Xie, Phys. Rev. B \textbf{80}, 165316 (2009).

\bibitem{Furd1988} J. K. Furdyna, J. Appl. Phys. {\bf 64}, R29 (1988).
\bibitem{Beck2001} C. R. Becker et al., Phys. Stat. Sol. (b) {\bf 229}, 775 (2001).

\bibitem{Liu} C.-X. Liu et al., Phys. Rev. Lett. \textbf{101}, 146802 (2008).
\bibitem{Liu2013} X. Liu, H.-C. Hsu, and C.-X. Liu, Phys. Rev. Lett. {\bf 111}, 086802 (2013).

\bibitem{Datta} S. Datta, \emph{Electronic Transport in Mesoscopic Systems} (Cambridge University Press, Cambridge, UK, 2003).
\bibitem{Ando} T. Ando, Phys. Rev. B \textbf{44}, 8017 (1991).


\bibitem{Zhou2008} B. Zhou, H.-Z. Lu, R.-L. Chu, S.-Q. Shen, and Q. Niu, Phys. Rev. Lett. {\bf 101}, 246807 (2008).

\bibitem{Datt1990} S. Datta and B. Das, Appl. Phys. Lett. {\bf 56}, 665 (1990).
\bibitem{Zuti2004} I. Zutic et al., Rev. Mod. Phys. {\bf 76}, 323 (2004).

\bibitem{Roth2010} D. G. Rothe et al., New J. Phys. {\bf 12}, 065102 (2010).

\bibitem{Takagaki} Y. Takagaki, Phys. Rev. B \textbf{85}, 155308 (2012).




\end{thebibliography}
\end{document}